\documentstyle[11pt,newpasp,twoside]{article}
\def\edcomment#1{\iffalse\marginpar{\raggedright\sl#1\/}\else\relax\fi}
\reversemarginpar

\begin{document}
\title{Radio Halos and Relics in Clusters of Galaxies and Detection Statistics}

\author{Gabriele Giovannini}

\affil{Dipartimento di Fisica, viale B. Pichat 6/2,
40127 Bologna, Italy \\
and Istituto di Radioastronomia, CNR, Bologna, Italy}

\author{Luigina Feretti}
\affil{Istituto di Radioastronomia, CNR, via Gobetti 101, 40129 Bologna, Italy}

\author{Federica Govoni}
\affil{Dipartimento di Astronomia, via Ranzani 1, 40127 Bologna, Italy}

\begin{abstract}
New sensitive VLA observations confirm the existence of halo and
relic sources in 6 Abell clusters
where a diffuse emission was found in the NVSS.

We find evidence that the frequency of clusters with halos and relics
is larger in clusters with high X-ray luminosity.
The evidence that the occurrence of a halo source is larger at high redshifts is
marginal.

\end{abstract}

\section{VLA observations}

In a recent paper, Giovannini, Tordi, \& Feretti (1999) presented a sample of 
halo and relic cluster radio sources found in the NRAO VLA Sky Survey (NVSS,
Condon et al., 1998). They used as cluster sample the X-Ray-brightest 
Abell-type clusters (XBACs)
presented by Ebeling et al. (1996). The complete sample contains 177 clusters;
some more clusters are present but do not belong to the 
complete sample because of their too low galactic latitude or too high redshift
(larger than 0.2; see Giovannini, Tordi, \& Feretti (1999) for a detailed 
discussion).

Six clusters where a central halo and/or a peripheral relic source was found
for the first time, were observed by us 
with $\sim$ 5 hrs long observations each, with the VLA in the C and D 
configurations.
They are: A115, A520, A773, A1664, A2254 and A2744.

In all clusters
the presence of the extended source was confirmed. 
The new results are in good agreement with NVSS data, 
confirming the very high reliability of the NVSS survey also for low
brightness extended sources.
The size of diffuse sources is larger in the new maps, as expected because of 
the better sensitivity reached in pointed observations.

Adding to the list given by Giovannini, Tordi, \& Feretti (1999)
all clusters where the presence of a halo and/or a relic source
was reported in literature, we have a sample of
$\sim$ 40 cluster of galaxies with a diffuse source.

\section{Frequency of halos and relics with the X-Ray luminosity}

The number of new diffuse halos detected in the NVSS is rather high, especially
considering that these sources are characterized by a steep spectrum and 
therefore that they are better imaged at frequencies lower than 1.4 GHz and
also taking into account the limited surface brightness sensitivity.

We used
these data to derive the occurrence of a diffuse source in clusters of galaxies
of different X-ray luminosity.
In Table 1 we report the results obtained for the complete sample (Col. 4)
and the total sample (Col. 5). The X-ray luminosity is computed in the Rosat
band (0.1 - 2.4 Kev).

\begin{table}
\begin{center}
\caption{Diffuse sources in clusters with different L$_x$}
\medskip
\begin{tabular}{ccccc}
\tableline
L$_x$ range      &  Cluster & Source    & \% & \% adding the \\
erg s$^{-1}$     &  number  &  number &    & not complete sample   \\
\hline
L$_x$ $\le$ 3$\times 10^{44}$ &     79           &     2        &   3 & 2 \\
3$\times 10^{44}$ $<$ L$_x$ $\le$ 5$\times 10^{44}$ & 34    &   2   & 6 & 6 \\
5$\times 10^{44}$ $<$ L$_x$ $\le$ 7$\times 10^{44}$ &  22    &   2 &  9 & 9 \\
7$\times 10^{44}$ $<$ L$_x$ $\le$ 10$^{45}$          & 31  &    9  & 29 & 27 \\
L$_x$ $>$ 10$^{45}$ & 11 & 4 & 36 & 44 \\
\tableline \tableline

\end{tabular}
\medskip \\
\end{center}
\end{table}

The percentage of clusters with halos and relics increases with X-Ray 
luminosity. There is evidence that this increase is more relevant for central 
halo sources than for peripheral relic sources.
The clusters hosting radio halos and/or relics
have an X-Ray luminosity significantly higher than the other clusters 
(confidence level $>$ 99.9\% estimated with a Kolmogorov-Smirnov test).

\section{Frequency of halos and relics with the redshift}

To investigate if the frequency of diffuse sources depends on the redshift, 
we have considered
only the 19 clusters with an X-ray luminosity $>$ 9 $\times$ 10$^{44}$ 
erg s$^{-1}$
which are always visible in the complete sample (z $\le$ 0.2). 
We find that 55\% of clusters with 0.15 $<$ z $\le$ 0.2 have a diffuse source,
to be compared with 33\% of clusters with 0.044 $\le$ z $\le$ 0.1. 
The difference is not statistically significant because the too small
number of clusters; a 
deeper and larger sample is necessary to investigate this point.

\end{document}